\documentclass[twocolumn,prl,a4paper]{revtex4}

\usepackage[T1]{fontenc} 
\usepackage[applemac]{inputenc}
\usepackage{graphicx}
\usepackage{dcolumn}
\usepackage{bm}
\usepackage{amsmath,amssymb}
\usepackage{pifont}
\usepackage{graphicx}
\usepackage{multibox}
\usepackage{epic}
\usepackage{ulem}

\usepackage{epstopdf}
\usepackage[usenames,dvipsnames]{color}

\def\rayer#1{\sout{}}

\begin{document}


\title{Surface Nanobubble Nucleation Visualized with TIRF Microscopy}

\author{Chon U Chan \& Claus-Dieter Ohl}

\affiliation{Division of Physics and Applied Physics, School of Physical and Mathematical Sciences, Nanyang Technological University, Singapore 637371, Singapore}
\date{\today}

\begin{abstract}
Nanobubbles are observed with optical microscopy using the total internal reflection fluorescence (TIRF) excitation. We report on TIRF visualization using Rhodamine 6G at 5$\mu\,$M concentration which results to strongly contrasted pictures. The preferential absorption and the high spatial resolution allow to detect nanobubbles with diameters of 230\,nm and above. We present a study of the nucleation dynamics from the water-ethanol-water exchange and report the size distributions. Nanobubble nucleation is observed within 4 min after the exchange, later a stable population of nanobubbles with a surface density of 0.55 bubbles\,/$\mu$m$^2$ is formed. Interestingly, unstable, slowly dissolving nanobubbles are observed during the first stage of water-ethanol exchange; only after the ethanol-water exchange stable nanobubbles appear.
\end{abstract}

\maketitle

\section{Introduction}

Nanobubbles are nanometer high gas bodies attached to a surface being immersed in a liquid \cite{SeddonLohse2011}. In particular the study of surface stabilized nanobubbles in water has attracted a lot of attention because of their potential role for interfacial water technologies~\cite{Liu2009,Wang2010}. So far nanobubbles are mostly studied with scanning atomic force microscopy (AFM) which offers a very high spatial resolution (below 1 nm). The downside is the long scanning time which prevents a study of their short time dynamics. Higher temporal resolution has been achieved with single line scans of the AFM tip \cite{Yang2009} or with IR spectroscopy, e.g.~\cite{Zhang2007,Zhang2008}. 

In this Letter we report on the visualization of stabilized nanobubbles on a {\em hydrophilic} surface with a standard optical microscopy technique. 
A common procedure to create surface nanobubbles on hydrophilic interfaces is the water-ethanol-water exchange process. There, the water is replaced first by ethanol and then by water again~\cite{Lou2002}. The higher solubility of gas in ethanol and the exothermic mixing of ethanol with water leads to the release of dissolved gas from the ethanol. The subsequent replacement of the ethanol with water is made responsible for the nucleation  nanobubbles from the now supersaturated water, see~\cite{Lou2000} and further studies summarized in Ref.~\cite{SeddonLohse2011}.
     
The recent review by Seddon and Lohse~\cite{SeddonLohse2011} pointed out that for success in understanding the physics of nanobubbles the experimental reproducibility has to be improved, the nucleation process detailed, and indications on a possible dynamic equilibrium collected. All these questions demand for an experimental technique which not only can resolve the surface nanobubbles but also study their dynamics with a much better temporal resolution as currently available. In this Letter we present an experimental technique which can help to solve some of the remaining puzzles in nanobubble research.

We first describe the experimental technique to visualize surface nanobubbles and then discuss the dynamics of bubble formation during the water-ethanol-water exchange using this optical technique. The physical principle behind optical nanobubble detection is total internal reflection fluorescence (TIRF) microscopy. TIRF allows to illuminate only a very thin volume of liquid in contact with an interface, i.e. between glass and water.  For our geometry the total internal reflection occurs for an angle $\theta_{tot} > \sin^{-1}(n_w/n_g) \approx 61^\circ$, where $n_w=1.33$ and $n_g=1.52$ are the indexes of refraction for water and the microscope cover glass, respectively. In the experiment we can achieve conveniently this angle of incidence; the maximum possible angel $\theta$ has been determined with a prism and is about $74^\circ$. The corresponding penetration depth (that is where the intensity drops to 1/e) can be estimated~\cite{Axelrod1984} as 70nm by $z_{0}=\lambda(4\pi\sqrt{(n_{g}sin\theta)^{2}-n_{w}^{2}})^{-1}$, where $\lambda$ is the laser wavelength.

\begin{figure}[h]
\centerline{\includegraphics[width=3.3in]{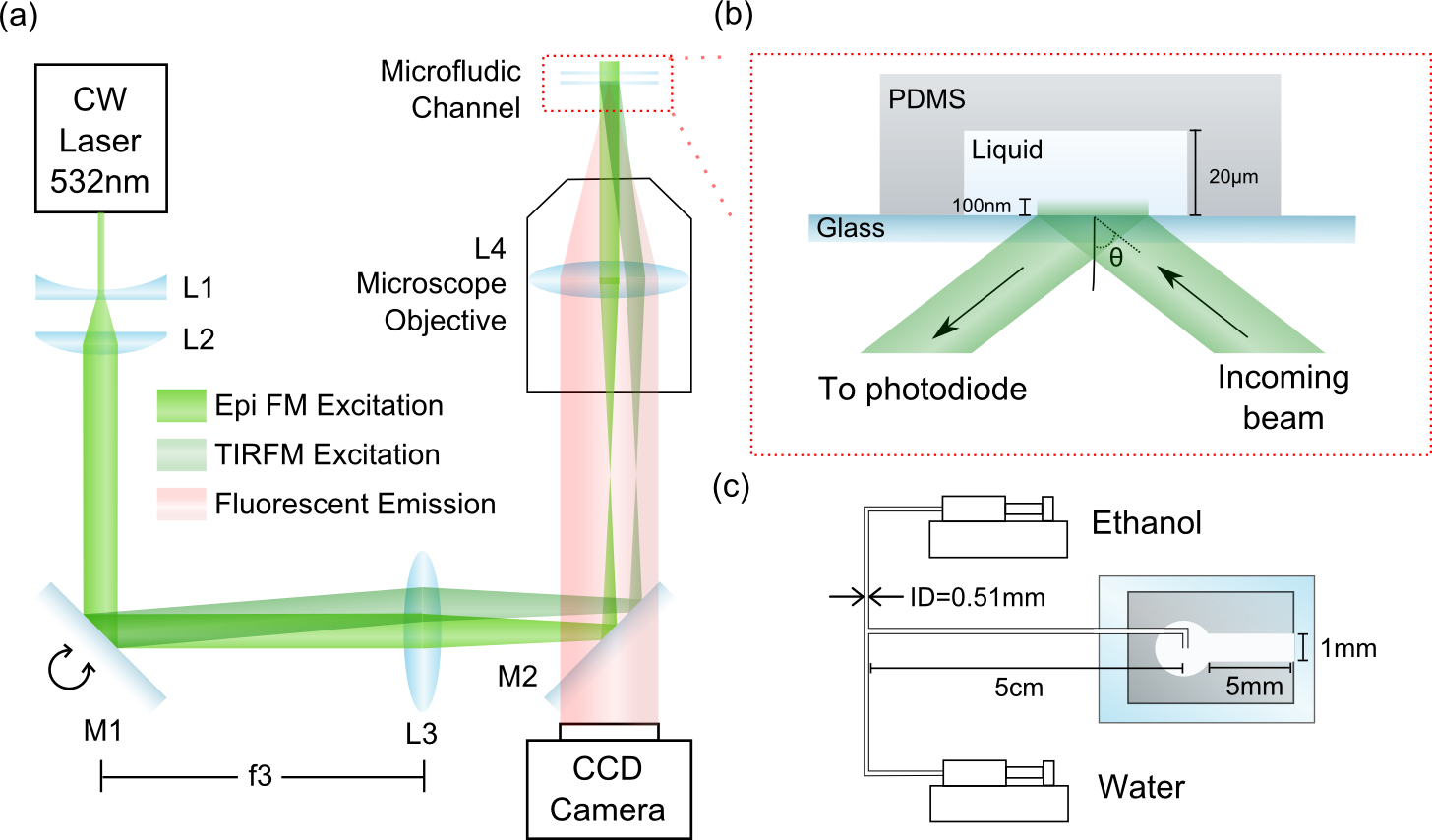}}
\caption{\label{setup}
(a) Sketch of the optical path for TIRF microscopy. Epifluorescence and TIRF microscopy excitations are available by rotating mirror M1. (b) Cross session of the PDMS channel to study the water-ethanol-water exchange. (c) Schematic of the flow line consisting of two syringes and syringe pumps, a T-connection, and a 5 cm feed to the microchannel. The microchannel is 5 mm long and 1 mm wide.} 
\end{figure}

\section{Experimental Setup}

Figure~\ref{setup}a depicts the beam forming and injection into the microscope. We use a green DPSS CW laser (8mW, $\lambda = 532\,$nm) expanded to 12 mm in diameter and steered into the side port of an inverted microscope (Olympus IX71). A mirror and a lens set the angle of incidence $\theta$, the mirror (M1) steers and lens (L3) focuses the beam into the back focal aperture of the microscope objective (Olympus ApoN 60x, NA 1.49). By adjusting the rotating mirror M1 located at the back focal plane of L3, the incident angle $\theta$ can be varied while the illumination spot remains fixed and within the field of view. Due to the limited working distance of the high-NA microscope objective only thin glass plates (here a cover slip glass \#1 with $140\,\mu$m thickness) can be used. A cooled slow scan CCD camera (Sensicam QE, PCO, Germany) with a pixel size of $6.45\,\times\,6.45\,\mu$m$^2$ is used for imaging. We have measured a pixel resolution of 108 nm/pixel. The diffraction limited resolution for the objective is about 230\,nm (at 550\,nm).  The camera records a field of view of $55\,\times\,69\,\mu$m$^2$ with a typical exposure time of 25-40\,ms at 18 frames/s. This is sufficiently fast to resolve the nanobubble nucleation dynamics (see below).

Figure~\ref{setup}b sketches the side view of the beam reflecting from the glass-liquid interface. The liquid is transported in a microchannel with rectangular cross-section (1 mm width, 20$\,\mu$m height). The channel is fabricated with standard soft lithography technique, the patterned PDMS channel bonded on a glass cover slip (Menzel-Glaser, Germany) pre-cleaned in an ultrasonic bath. $5\,\mu$M Rhodamine 6G fluorescent dye is dissolved both in the DI water (purification with Sartorius Arium 611vf, France) and in 99\% ethanol (Riverbank Chemicals, Singapore). Both liquids are loaded into separate syringes. They are discharged in sequence using syringe pumps. Before the liquids reach the channel they flow through a T-junction which is about 5\,cm away from the channel inlet. When the flow channel is operated some pressure builds up which eventually flexes the cover slip glass mildly. Therefore the distance of the microscope objective has to be adjusted during the experiment. The manual height adjustment is aided by a back reflection of the laser beam. This reflection  passes over an aperture and is detected with a photodiode (not shown in Fig.~\ref{setup}a). During the  experiments the photodiode signal is kept constant by adjusting the stage manually and thus keeping the distance betwene objective and glass constant. 

\section{Results}

\begin{figure}[h]
\centerline{\includegraphics[width=3.3in]{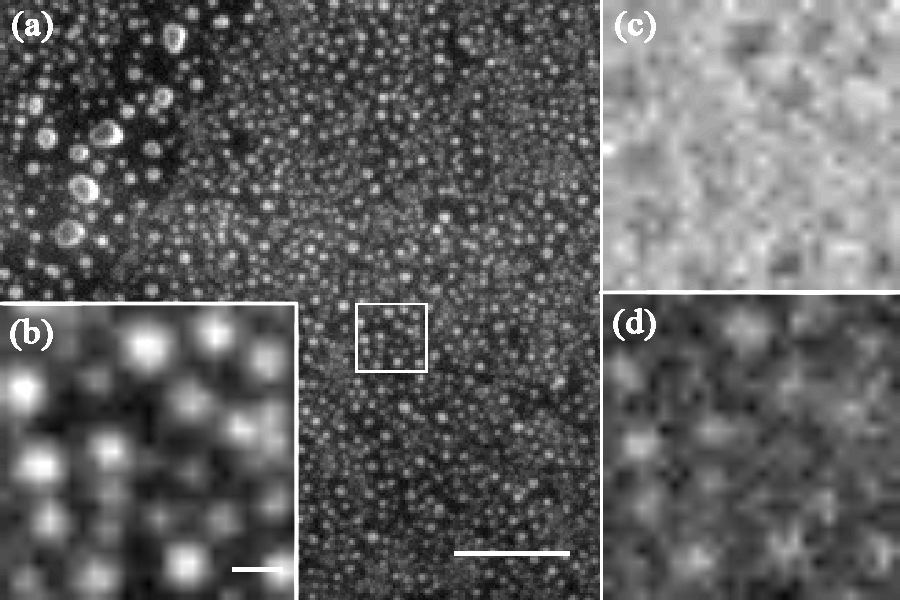}}
\caption{\label{overview}
(a) Nanobubbles observed under TIRF microscopy. Scale bar is $5\,\mu$m. The square area is zoomed into for comparison of the different technique (b) TIRF microscopy (scale bar is 500nm), (c) brightfield and (d) epifluorescence.} 
\end{figure}

Figure~\ref{overview}a demonstrates a typical figure recorded after the water-ethanol-water exchange in TIRF mode. The surface lights up with disc shaped objects filling almost completely the surface of the glass cover slip. Before the exchange an unstructured dim light is recorded. Yet, after the exchange protocol strong contrasted objects appear. We attribute these objects with surface nanobubbles. The typical contrast of the surface bubbles as defined by $(I_{max}-I_{min})/(I_{max}+I_{min})$ is approx. 0.5 and the signal to noise ratio is 3. 

Figure~\ref{overview}b-d are enlarged views of the area indicated with a white square in Fig.~\ref{overview}a (scale bar length is 500 nm). For magnification they are re-sampled using linear splines. Figure~\ref{overview}b is just the enlarged part of Fig.~\ref{overview}a, while Fig.~\ref{overview}c and Fig.~\ref{overview}d compare the same view now in brightfield mode and in epifluorescence ($\theta = 0^\circ$), respectively. From this comparison it becomes clear that the TIRF microscopy provides a strong contrast for nanobubble visualization. Nanobubbles are not visible in the brightfield mode, and only some of the largest structures appear but weak in the epifluorescence mode. 

What leads to the stark contrast of nanobubbles under TIR illumination? It is well known that Rhodamine is accumulating at liquid-gas interfaces. Zheng {\it et al.}~\cite{Zheng2001} demonstrate that the fluorescence is about 60 times stronger from the water-air interfaces as compared to the bulk. Thus we explain the strong contrast of nanobubbles in TIRF microscopy by the combination of adsorption of Rhodamine at the nanobubble interface and the short penetration depth of the evanescent wave exciting preferentially the liquid on the scale of the nanobubble height, i.e. about 100\,nm.

\begin{figure}[h]
\centerline{\includegraphics[width=3.3in]{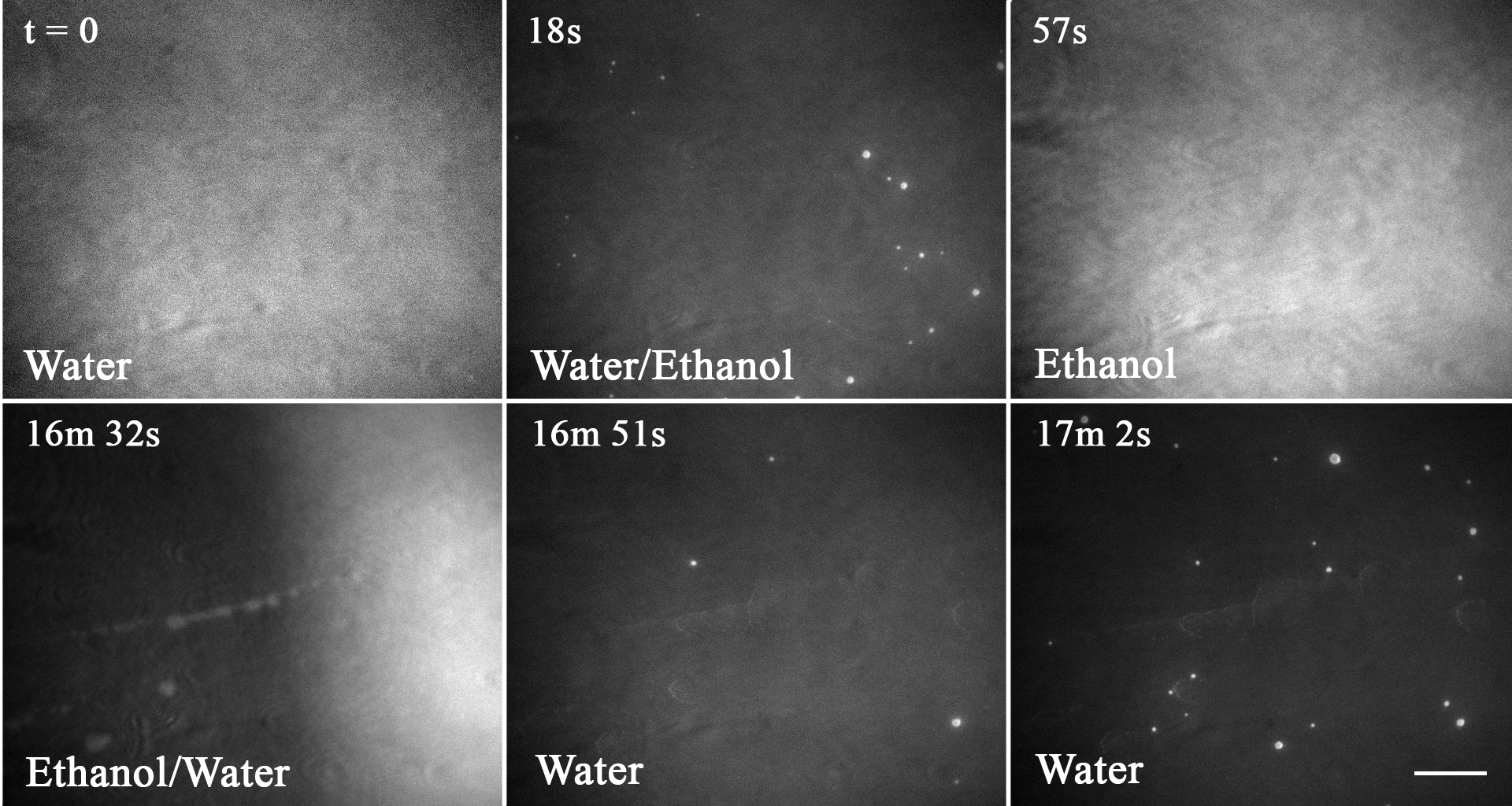}}
\caption{\label{exchange} Snapshots of images obtained during water-ethanol-water exchange. The intensity is normalized for each frame. Upper row: Bubbles are formed when water is replaced by ethanol. These bubbles dissolved quickly. Lower row: Bubbles again nucleate when ethanol is replaced by water. Bubbles continue to nucleate after the exchange is finished. Scale bar is $10\,\mu$m.} 
\end{figure}

Although the water-ethanol-water exchange is a common method to generate nanobubbles on hydrophilic surfaces, very few measurements on the nucleation dynamics are available. The microbalance techniques predicts that the nanobubbles are formed within one minute after the liquid exchange~\cite{Zhang2008b}.

Figure~\ref{exchange} presents the different stages leading to a surface decorated with nanobubbles. For this experiment, both syringes holding water and ethanol are discharged one after another. Initially the feed line and the whole microchannel, see Fig.~\ref{setup}c, is filled with water. Then a fast flow (Ethanol) of 125$\mu\,$l/min transports the water-ethanol interface from the T-juntion to the inlet of the microchannel. Thereby we reduce diffusion at the T-junction liquid-liquid interface. Then a slow flow of 1$\mu\,$l/min pushes the interface through the microchannel. The dynamics is presented in Fig.~\ref{exchange}: at time $t = 0$ the channel is filled with water only. At time $t = 18\,$s the water is partly replaced from the left with ethanol and quickly bright spots appear on the surface. These nanobubble however quickly dissolve and after 40\,s all bubbles have dissolved. Then, we keep ethanol for about 15 minutes in the channel before starting the replacement. Again, for this exchange the water is pushed initially with a fast flow through the feed line followed by a slow flow through the microchannel. At $t = 16\,$min$\,32\,$s the water front arrives at the field of view and pushes out the ethanol to the right. At time $t = 16\,$min$\,51\,$s the microchannel is completely filled with water and the  nanobubbles start to nucleate. 11 s later many more nanonbubbles have nucleated, see last frame of Fig.~\ref{exchange}. 

\begin{figure}[h]
\centerline{\includegraphics[width=3.3in]{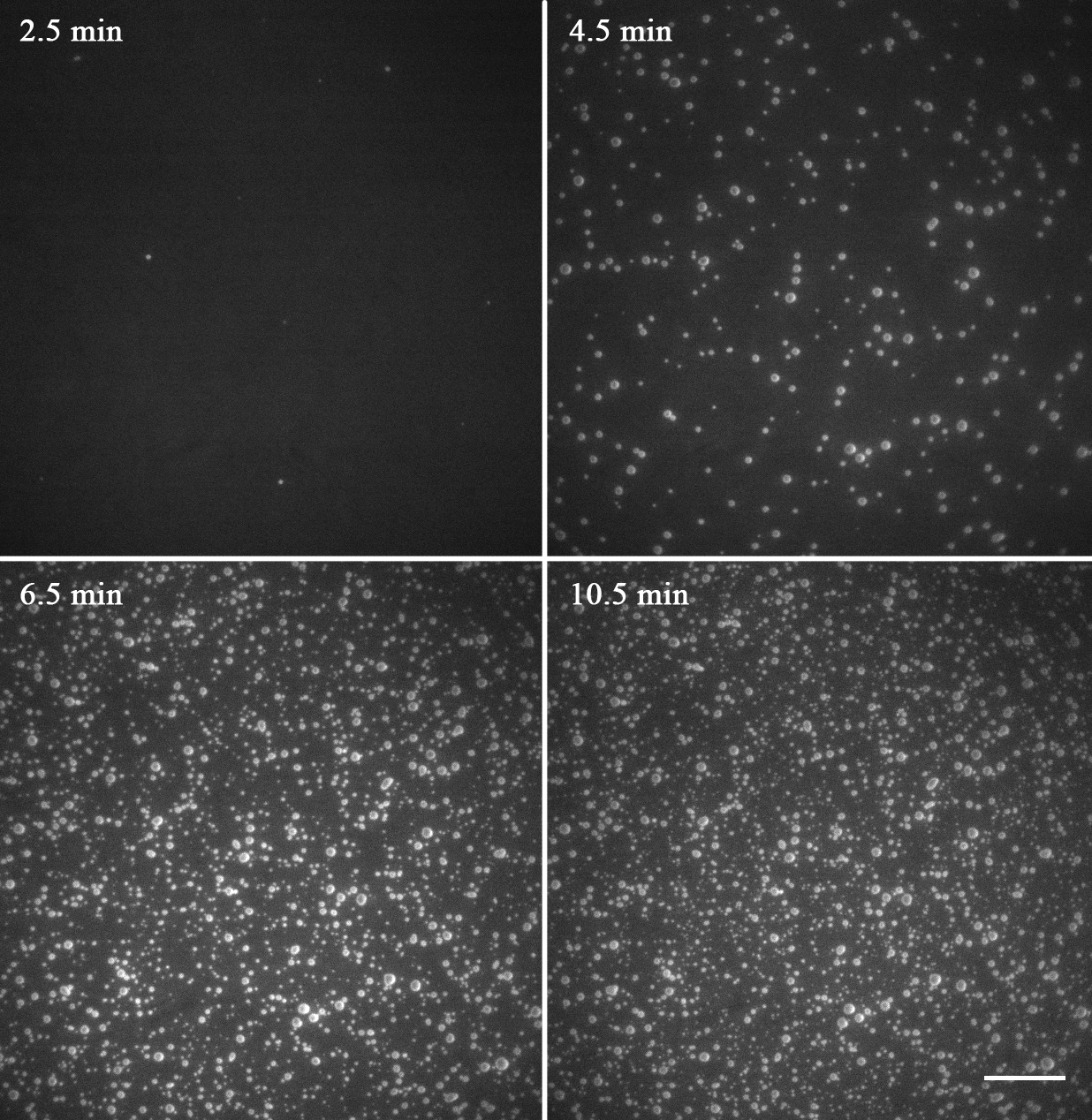}}
\caption{\label{after_exchange} The channel is filled with water after water-ethanol-water exchange. Bubbles nucleate gradually over several minutes. The exchange flow is from left to right. The Scale bar is $10\,\mu$m.} 
\end{figure}

Experiments not detailed here revealed that the flow rate affects the nucleation speed. At higher flow rates (about 50$\mu\,$l/min) nucleation completes almost instantly once the liquid-liquid interface has passed. While at smaller flow rates, e.g. 5$\mu\,$l/min and below, nucleation may take several minutes before a steady nanobubble population is reached. 
The temporal development of nanobubble nucleation at a flow rate of 5$\mu\,$l/min is shown in Fig.~\ref{after_exchange}. At time $t = 2.5\,$min after water-ethanol-water exchange a few isolated nanobubbles appear. 2 minutes later the surface shows about 0.1\,bubbles\,/$\mu$m$^2$; another 2 minutes the surface density has quadrupled. Figure 5a plots the nanobubble density as a function of time. The rise time of the nanobubble density, i.e. from 10\% to 90\% of the maximum bubble density, is about 240\,s. An analysis of the nanobubble diameter is presented in Fig. 5b. We understand that bubbles below the resolution limit of about 230nm can't be resolved. Thus, the distribution only shows relatively large diameter nanobubbles. Similar sizes with diameters of up to $1\,\mu$m have been observed on rough surfaces~\cite{Borkent2010}.

\begin{figure}[h]
\centerline{\includegraphics[width=3.3in]{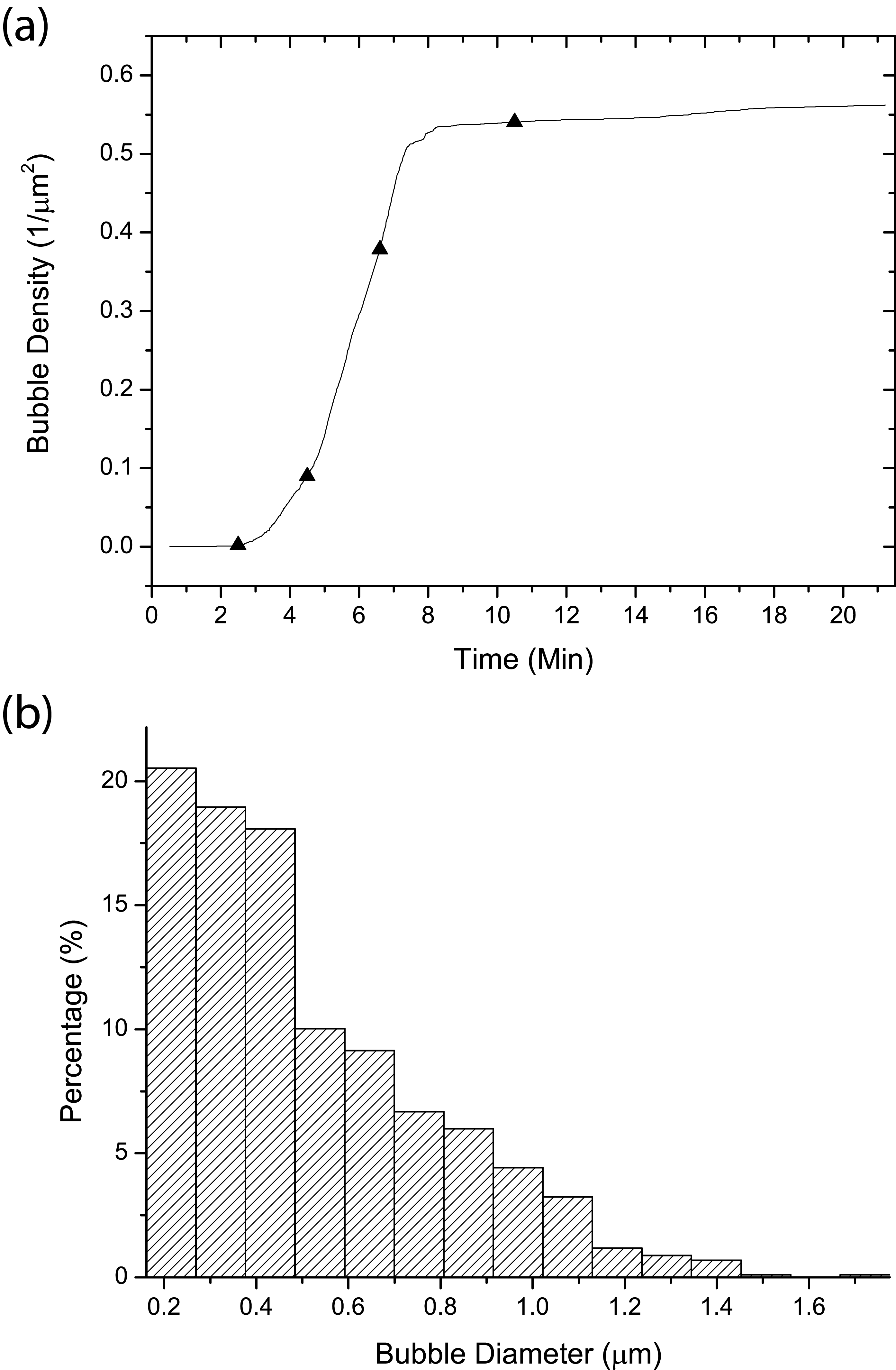}}
\caption{\label{plots} A) The bubble density plotted as a function of time for the experiment shown in Fig.~\ref{after_exchange}. The triangles state the times where the snapshots in Fig.~\ref{after_exchange} are taken. B) Size distribution of bubbles formed. Bins increase by the pixel size. The smallest bin is 2 pixels wide.} 
\end{figure}

In summary we have presented the formation of nanobubbles on a hydrophilic glass surface during the water-ethanol-water exchange in a microchannel. The growth dynamic has been captured optically with a TIRF microscope. We achieved a temporal resolution with a cooled CCD camera and a low power green laser of 56 ms. The pixel resolution of our setup allows observing nanobubbles with diameters of 230nm and above.
Having this technique available may help to address some of the open questions on interfacial nanobubbles. A non-exhaustive list of interesting experiments that now become possibles is the study of superstability to strong tension waves~\cite{Borkent2007}, acoustic resonances~\cite{Rathgen2007} of randomly distributed nanobubbles, drag reduction~\cite{Steinberger2007,Wang2010} in nanochannels and their stability to shear flow, Brownian motion near nanobubbles, and the study of diffusional growth of gas bubbles from nanobubbles.  

\acknowledgements
We thank Lam Research AG in particular Frank Holsteyns and Alexander Lippert for their continuous support, Lin Ma for helpful discussions on dyes.

\end{document}